\documentstyle[prb,aps,preprint,floats]{revtex}
\begin{document}
\hsize\textwidth\columnwidth\hsize\csname@twocolumnfalse\endcsname
\draft  
\title{A study  of the superconducting gap in $R$Ni$_2$B$_2$C ($R$ = Y, Lu) 
single crystals by  inelastic light scattering}
\author{
In-Sang Yang$^{1,2}$,
M. V. Klein$^2$, S. L. Cooper$^2$,
P. C. Canfield$^3$,
B. K. Cho$^{4,*}$, Sung-Ik Lee$^4$
}
\address{$^1$Department of Physics, Ewha Womans University, Seoul 120-750,  
Korea\\
                 $^2$Materials Research Laboratory, Department of Physics, 
                        University of Illinois at Urbana-Champaign, Urbana, IL 61801,
                 $^3$Ames Laboratory, Department of Physics and Astronomy, 
                        Iowa State University, Ames, IA 50011,
                 $^4$National Creative Research Initiative Center for 
Superconductivity, 
                       Department of Physics, Pohang University of Science and 
Technology, 
                       Pohang 790-390, Korea
                 }
\date{\today}
\maketitle
\begin{abstract}
Superconductivity-induced changes in the electronic Raman scattering response 
were observed
for the  $R$Ni$_2$B$_2$C ($R$ = Y, Lu) system in different scattering geometries.
In the  superconducting state, 2$\Delta$-like peaks were observed in A$_{1g}$, 
B$_{1g}$, and 
B$_{2g}$ spectra from single crystals. 
The peaks in A$_{1g}$ and B$_{2g}$ symmetries are significantly sharper and 
stronger than the 
peak in B$_{1g}$ symmetry.
The temperature dependence of the frequencies  of the 2$\Delta$-like peaks shows
 typical BCS-type behavior,  but
the apparent values of the $2\Delta$ gap are strongly anisotropic for both systems. 
In addition, for both YNi$_2$B$_2$C and LuNi$_2$B$_2$C  systems, 
there exists reproducible scattering strength below the  $2\Delta$ gap 
which is roughly linear to the frequency in B$_{1g}$ and  B$_{2g}$ symmetries.
This discovery of scattering below the gap in non-magnetic borocarbide 
superconductors, 
which are thought to be conventional BCS-type superconductors, 
is a challenge for current understanding of superconductivity in this system.
\end{abstract}
\pacs{PACS numbers: 78.30.Er;  74.70.Dd;  74.25.Jb  
Key Words: borocarbide superconductors; superconducting gap; electronic Raman 
scattering } 

\section{Introduction}
The rare-earth borocarbides are believed  to be 
conventional BCS-type superconductors\cite{hoellwarth}
with a  number of interesting properties that are not fully understood.
Borocarbides  are similar to intermetallic 
superconductors such as V$_3$Si and Nb$_3$Sn
in that they have a relatively strong electron-phonon coupling constant 
and a moderately large density of states at the Fermi level.\cite{fermi}
While it is generally agreed that the electron-phonon interaction is the underlying 
mechanism
for superconductivity in these systems,  the responsible phonons are claimed to 
be either
a high-energy phonon (boron A$_{1g}$ near 850 cm$^{-1}$) \cite{mattheiss}
or a low-energy soft-mode phonon.\cite{yanson}
Additionally, the borocarbides have  relatively simple  tetragonal crystal 
structure,
yet they show a variety of physical properties that have  been active research 
topics over the 
past years.
For example, some of the borocarbides exhibit magnetic ordering  and 
superconductivity 
at around the same temperature, 
offering the possibility of studying the interplay between  superconductivity 
and magnetism.\cite{canfield}

Even the non-magnetic rare-earth borocarbides  $R$Ni$_2$B$_2$C ($R$  = Y, Lu)  
show rich superconducting phenomena, 
such as anisotropic upper critical fields 
and angular dependence of in-plane magnetization,\cite{nonlocal}
a hexagonal-to-square vortex lattice transition,\cite{paul,eskildsen,gammel}
and phonon-softening at a finite wave vector 
where strong Fermi-surface nesting is suggested.\cite{INS}
The results of inelastic neutron scattering measurements\cite{INS} are on soft 
$q \neq 0$ phonons, 
which would not be detected  using Raman spectroscopy.

A study of the low-energy excitation spectra would  clearly lead to a more complete 
understanding of the
electronic interaction in these systems.
The superconducting gap has been a  major subject of research in superconductivity.
Up to now, the gap in borocarbides has been studied by  tunneling 
spectroscopy\cite{tunneling}  and infra-red (IR) reflectivity\cite{bommeli} 
which are insensitive to the wave vector-space dependence of the gap.
Electronic Raman spectroscopy is sensitive to the wave vector dependence of the 
gap.  
When the optical penetration depth exceeds the superconducting coherence length, 
it yields a susceptibility for two quasi-particles of essentially zero total wave 
vector.  
The quasi-particles that contribute to the response are those with wave vectors 
close to the Fermi surface 
for which the Raman vertex has a large modulus squared.  
This vertex equals the projection onto the polarization vectors of the incident 
and scattered 
photons of a tensor closely related to the inverse effective mass tensor.  
Manipulation of the polarization vectors changes the Raman symmetry and selectively 
probes 
the gap near certain regions of the Fermi surface.\cite{vertex}
Raman spectroscopy studies have played an important role in understanding 
superconductivity
of conventional\cite{a15} and exotic superconductors.\cite{cooper} 
For example, superconductivity-induced redistribution of the electronic continua
of the high-T$_c$ superconductors led  to important clues as to the  d-wave nature 
of the 
order parameter.
However, there remain  unsolved puzzles regarding  different values of   the 
2$\Delta$  peak 
positions in different scattering geometries, 
the power-law behavior  of the scattering response below 
2$\Delta$ and its relation to the d-wave nature of the order parameter, and to 
the role of
impurities.

Until now there has been only one kind of ``conventional'' superconductors ($A15$) 
whose gap was
measured by electronic Raman, while there are numerous oxide superconductors 
measured by the same technique. In other words, we do not really know how the 
conventional superconductors behave in the electronic Raman studies.
Without a sound background for comparison, many of  the studies of 
``unconventional'' behavior of the exotic superconductors would be meaningless. 
From number of aspects (including our own measurements), borocarbides seem to 
behave 
conventionally. Thus borocarbides would be suitable choice of superconductor 
for  investigation usingelectronic Raman spectroscopy.

In this paper, we  report an observation  of clear redistribution  of the 
electronic  continua in 
different  scattering  geometries  in  the   $R$Ni$_2$B$_2$C   ($R$ =  Y,  Lu)    
borocarbides  in 
the superconducting state.
2$\Delta$-like peaks were observed in the  superconducting state in A$_{1g}$, 
B$_{1g}$, and 
B$_{2g}$ Raman symmetry from single crystals. 
The peaks in A$_{1g}$ and B$_{2g}$ symmetries are significantly sharper and 
stronger than the 
peak in B$_{1g}$ symmetry.
The temperature dependence of the frequencies of the peaks shows
typical BCS-type behavior of the superconducting gap, but
the apparent values of the $2\Delta$ gap are strongly anisotropic for both systems. 
In contrast to what we normally expect from conventional superconductors, 
we find reproducible scattering strength below the  2$\Delta$-like peak in all 
the scattering
geometries. 
This is the first Raman measurement  that shows finite scattering below the 
gap-like 
peaks of the ``conventional'' superconductors.
Previous measurements on $A15's$ were  too noisy to detect any finite scattering 
below  the gap.\cite{a15}

\section{Experiment}

All the samples measured in this work were single crystals grown by the flux-
growth 
method.\cite{canfield,xu}
Samples  were   characterized by   various  measurements,   including 
resistivity   versus 
temperature, magnetization versus temperature, and neutron 
scattering.\cite{character}
The crystal structure of   $R$Ni$_2$B$_2$C is tetragonal body-centered space group 
$I4/mmm$, 
and phononic Raman analyses have been made earlier.\cite{p-raman} 
Raman spectra were obtained 
in a pseudo-backscattering geometry 
using a custom-made subtractive triple-grating spectrometer
designed for very small Raman shifts and ultra low intensities.\cite{kang} 
3 mW of 6471 \AA  \/  Kr-ion laser light was focused onto a spot of
$100 \times 100$ $\mu$m$^2$, which results in heating of the spot above the
temperature of the sample environment.
The temperature of the spot on the sample surface  
was estimated by solving the heat-diffusion equation.
$\Delta$T, the rise in temperature  is 
proportional to the inverse of the thermal conductivity of the sample.
$\Delta$T is largest  at lowest temperature  because the  thermal conductivity is  
smaller at 
lower temperatures.\cite{thermal} 
The estimated $\Delta$T is 2.7K at 4K and 0.9K at 14K for the case of
YNi$_2$B$_2$C single crystal.
The ambient temperature at which the Raman continua begin to show the  
redistribution was
determined to be 14K, which is in agreement with this estimation.
All the spectra have been corrected for the Bose factor. They therefore are 
proportional to the
imaginary part of the Raman susceptibility.

\section{Results and Discussion}

Superconductivity-induced changes in the B$_{2g}$ and B$_{1g}$ spectra of 
YNi$_2$B$_2$C taken at different  temperatures are shown in Fig.~1.  
The quality of the sample surfaces was good enough that
Raman spectra could be taken down to $\sim$10 cm$^{-1}$.
Sharp drops of the spectra below 10  cm$^{-1}$ (for B$_{2g}$ and B$_{1g}$) or 20  
cm$^{-1}$ 
(for A$_{1g}$) are due to cutoff of the filtering stage, 
and the sharp rises near zero frequency are due to the laser line (6471 \AA ).
The relative strengths of the spectra in different scattering geometries 
 are meaningful although the absolute values of the scattering
strength are in arbitrary units in all the figures.
In the superconducting state (T $\approx$  7 K), the spectra show  a  clear 
redistribution of 
the scattering weight: depletion of the weight at low frequencies ($\leq$ 40 
cm$^{-1}$) and
the accumulation of the weight above $\sim$ 40 cm$^{-1}$, resulting in sharp
gap-like peaks in both B$_{2g}$ and B$_{1g}$ symmetries.
The gap-like feature is more prominent and sharper in B$_{2g}$ symmetry 
than in B$_{1g}$ symmetry. 
The sharpness of the B$_{2g}$ peak suggests that its origin may be different from 
that of B$_{1g}$.

The peak at 200 cm$^{-1}$ is the Ni B$_{1g}$ phonon mode, the height of which (1250 
on 
the same scale) is much stronger than the B$_{1g}$ electronic Raman peaks.
Previous Raman measurements on borocarbides  concentrated mainly on phononic 
peaks,\cite{p-raman}
and the gap features  were not observed in earlier  electronic Raman 
measurements.\cite{e-raman}
Electronic Raman spectra from borocarbides are so weak that a sensitive 
spectrometer 
with good stray-light rejection is required.

The peak frequency $\omega_p$, read from graphs,  of the peak feature 
taken at 7 K is normalized in such a way 
that its normalized value is the same as the value of the reduced 
superconducting gap $\Delta$(T=7K)/$\Delta_0$  (= 0.95) in the 
BCS theory, where $\Delta_0$ = $\Delta$(T = 0), at the reduced temperature of 0.45 
(= 7/15.5).
$\omega_p$(0), the values of the peak frequency in the limit of 0 K thus obtained 
are 40.1 cm$^{-1}$ for B$_{2g}$ and 48.9 cm$^{-1}$ for B$_{1g}$.
$\omega_p$/$\omega_p$(0), the normalized frequencies of the peaks of the peak 
features as
functions of the reduced temperatures  follow the curve for  
$\Delta$(T)/$\Delta_0$ predicted by 
the BCS theory as shown in the insets of the figure.
This suggests that  the peak features  arise due to  opening of the superconducting 
gap and provides some evidence that borocarbides are conventional BCS-type 
superconductors.

One of the advantages of the  borocarbide superconductors is that their
upper critical fields ($H_{c2} \leq 6 T$  at 6 K)  can be easily reached  by using 
commercial superconducting magnets.
We applied strong magnetic fields (up to 7 T) parallel to c-axis and investigated 
the effects 
on these gap-like features of both YNi$_2$B$_2$C and  LuNi$_2$B$_2$C crystals at 
low temperatures. 
The gap-like features were completely suppressed by the strong magnetic fields,
confirming that these features are indeed induced by superconductivity.
(See Fig.~2 a.)
Full analysis of the dependence on magnetic fields of the peak frequencies and 
the features below the gap will be presented in an additional manuscript.

Unlike  other   gap-probing  measurements  such   as  tunneling  experiments   
and  x-ray photoemission
spectroscopy (XPS), Raman spectroscopy is able to measure the
anisotropy of the gap at a high resolution of a few $meV$. 
In addition, electronic Raman spectroscopy has an advantage over 
infra-red spectroscopy\cite{bommeli} in studying the gap of
superconductors in their clean limit.
Figure~2 shows Raman spectra in  A$_{1g}$, B$_{1g}$, and B$_{2g}$ geometries  
of YNi$_2$B$_2$C and LuNi$_2$B$_2$C in superconducting
states and in normal states.  
Spectra in E$_g$ geometry were difficult to acquire from either fractured or 
polish-then-etched ac surfaces, even if those surfaces were optically flat. 
The A$_{1g}$ spectra reflect weighted average of 
the Fermi surface with a weighting function of $k_x^2 + k_y^2$ symmetry;
the B$_{1g}$  spectra, $k_x^2  - k_y^2$   symmetry; and the  B$_{2g}$ spectra,   
$k_xk_y$ 
symmetry.
The polarizations used were such that a contribution of A$_{2g}$ symmetry is 
possible in
all three cases; however, the A$_{2g}$ spectrum is expected to be negligible.
The most apparent phenomenon is the anisotropy in the intensity and sharpness of 
the 
gap-like features. 
The A$_{1g}$ and B$_{2g}$ peaks are significantly stronger and sharper than the 
B$_{1g}$ peak. Together with the smaller values of the positions of the 
A$_{1g}$ and B$_{2g}$ peaks than that of B$_{1g}$ peak as detailed below, this 
suggests that the sharp A$_{1g}$ and B$_{2g}$ peaks might be a reflection of 
collective modes. Further investigation is necessary to clarify the origin of the 
peaks.

The B$_{1g}$ gap energy is considerably larger
than the B$_{2g}$ gap energy signifying perhaps that the gap parameters are 
anisotropic for both 
YNi$_2$B$_2$C and LuNi$_2$B$_2$C. 
The apparent 
gap anisotropy is more pronounced in YNi$_2$B$_2$C than in LuNi$_2$B$_2$C.
The values extrapolated at T = 0 are 
 $\Delta_0$ (A$_{1g}$) = 2.5 meV (2$\Delta_0$/kT$_c$ = 3.7),
 $\Delta_0$ (B$_{1g}$) = 3.0 meV (2$\Delta_0$/kT$_c$ = 4.5), and
 $\Delta_0$ (B$_{2g}$) = 2.5 meV (2$\Delta_0$/kT$_c$ = 3.7) for YNi$_2$B$_2$C, 
while
 $\Delta_0$ (A$_{1g}$) = 3.0 meV (2$\Delta_0$/kT$_c$ = 4.4),
 $\Delta_0$ (B$_{1g}$) = 3.3 meV (2$\Delta_0$/kT$_c$ = 4.8),  and
 $\Delta_0$ (B$_{2g}$) = 3.1 meV (2$\Delta_0$/kT$_c$ = 4.5) for LuNi$_2$B$_2$C.
 These values are different from the values from other 
measurements.\cite{tunneling,bommeli}
 For instance, an IR reflectivity measurement reports
 $\Delta_0$ = 3.46 meV for polycrystalline  YNi$_2$B$_2$C and
 $\Delta_0$ = 2.68 meV for  single crystal LuNi$_2$B$_2$C.\cite{bommeli}

There is reproducible spectral weight below the $2\Delta$ peak in both symmetries 
for 
both YNi$_2$B$_2$C and LuNi$_2$B$_2$C  systems.
Special care  was taken in studying the features below the gap: Samples with 
high-quality surfaces were selected; higher resolution and 
longer exposure times were employed 
Figure~3 shows Raman spectra of YNi$_2$B$_2$C(B$^{10}$ isotope) in B$_{2g}$ and 
B$_{1g}$ symmetries. 
In this particular case, the cutoff of the filter state in the 
spectrometer was 6 cm$^{-1}$. 
As far as the spectral weight below the $2\Delta$ peak is concerned, 
samples with different boron isotopes give the same results.
Both B$_{2g}$ and 
B$_{1g}$ spectra show finite Raman scattering below the gap and it appears to be 
roughly linear to the Raman frequency below 30 cm$^{-1}$. 
For the B$_{2g}$ spectrum, there is an abrupt slope change at about 30 cm$^{-
1}$.
Similar linear-in-frequency  behavior was  observed in  the spectra  of 
LuNi$_2$B$_2$C  below 35 
cm$^{-1}$.
The possibility of instrumental broadening is ruled out by measuring the spectral 
widths of Kr-ion plasma
lines, which are about 2.5 cm$^{-1}$ in the spectral range of these measurements.

Such a feature below the gap could rise from different causes. 
We eliminate the possibilities of a) an extrinsic phase on the sample surfaces 
and b) inhomogeneity of the probing laser beam, and argue that 
this sub-gap feature is intrinsic to the borocarbide superconductors.
This would constitute the first, definitive  experimental observation of finite 
Raman scattering 
below the gap feature  for the ``conventional'' superconductors. 
Previous measurements on $A15$ compounds were simply too noisy to detect any finite 
scattering below 
the gap. \cite{a15}

a) The borocarbide single crystals do have extrinsic phases, mostly
from the flux. In our measurements, we were able to avoid such flux phases. 
If the sub-gap feature were due to an
extrinsic phase, the intensity of the sub-gap feature
would be dependent on the measurement conditions, e.g., dependent on
the fraction of flux under the exciting focussed laser spot.All the spectra
showed the same intensity, independent of where the spot was located. 
X-ray photoemission spectroscopy (XPS) analyses show that the flux
phase on the surface of the single crystals is Ni$_3$B or Ni$_3$BC.
Scanning electron microscopy (SEM) did not reveal any small-grained structures
on the surface or edges of the single crystals.
Auger analyses did not show any fine structures either.
Furthermore, they showed that chemical compositions did not vary significantly
from point to point on the surface.
All these analyses suggest that there is no fine-grained extrinsic phase 
contributing to the Raman spectra. 
In addition, we actually measured the Raman spectra of the
flux phase itself (Fig.~3 a).
Even if {\it all} the Raman intensity above
the gap feature is assumed to be due to the flux (over-estimation of the flux 
contribution), a finite Raman intensity 
below the peak feature is always obtained (under-estimation of the intrinsic 
contribution),
as shown in Fig.~3 b.

b) An experimental situation where 
part of the measured spot is normal (or close to the normal state) 
and another part is deep in the superconducting state due to inhomogeneity of the 
laser spot
could lead to finite scattering intensity below the gap.
This senario is excluded as follows:
1. In the measurements, spatial filtering  the excitation laser source was used 
to make the 
incident beam as homogeneous as possible.
2. We used several different laser intensities and confirmed the same feature 
below the gap.  The overall intensity scaled with the incident power
as long as the heating effect was small, so that the temperature of the measured 
spot 
remained well below T$_c$.
3. The sharpness of the B$_{2g}$ peak itself indicates that the actual temperature 
of the 
measured spot is quite uniform and well below  T$_c$.
Otherwise, we should not be able to see such a steep slope below the B$_{2g}$ peak.
Therefore, we believe that 
the finite Raman intensity below the peak feature is intrinsic.

 Cuprate superconductors show finite Raman response below the
2$\Delta$-like peak even in the superconducting state.\cite{cooper}
The lack of the complete suppression of the Raman response below the 2$\Delta$ 
peak
has been regarded  as  evidence that  the superconducting  gap function of  
cuprates has 
nodes where the gap vanishes.\cite{devereaux}
In simple BCS picture no density of states is allowed below the superconducting 
gap; thus, 
no Raman scattering beyond small smearing effects is expected below the gap in 
BCS-type 
superconductors.  
However, in this work, finite Raman scattering is observed below the gap of
borocarbides.
 
It is interesting to note that Kim\cite{kim} finds  finite  low-energy states below 
the gap in s-wave 
superconductors by considering a finite-ranged  pairing interaction energy. The 
quasiparticle density of states 
calculated numerically resembles the B$_{2g}$ Raman spectra shown in Fig.~3.
The underlying mechanism that gives  rise to the linear-in-frequency behavior 
below the 2$\Delta$ 
gap for the borocarbides, which are believed to be conventional BCS-type 
superconductors,
needs to be clarified further.

\section{Conclusion}

In conclusion,
sharp redistributions  of   the continuum  of  Raman spectra  were  observed  in  
different 
geometries
for the  $R$Ni$_2$B$_2$C ($R$ = Y, Lu) systems upon going into the 
superconducting states.
2$\Delta$-like features were observed in the A$_{1g}$,  B$_{1g}$, and B$_{2g}$ 
spectra of  the  $R$Ni$_2$B$_2$C ($R$= Y, Lu) single crystals in the 
superconducting states.
The peak features in  A$_{1g}$ and B$_{2g}$ geometry are sharper and stronger than
those in B$_{1g}$ geometry.
The  features are suppressed by applied magnetic fields, suggesting that
they are indeed closely related to the gap parameter 2$\Delta$.
The temperature dependence of the peaks of the 2$\Delta$-like features shows
typical BCS-type behavior of the gap values as a function of temperature.
However, the apparent  values of  the  gap  are anisotropic,  and the anisotropy  
is 
more pronounced in YNi$_2$B$_2$C than in LuNi$_2$B$_2$C.
Finite Raman scattering below the gap  was 
experimentally observed in these ``conventional'' superconductors. The finite 
scattering 
is intrinsic  and it appears to be linear in the Raman frequency. 
The lack of complete suppression of the Raman response below the 2$\Delta$ peak
of apparent  BCS-type superconductors needs to be further addressed. 

\acknowledgements
We thank  G. Blumberg, T. P. Devereaux, H.-L. Liu,  K.-S. Park, and M. R\"ubhausen 
for help and valuable discussions.
ISY and MVK were partially supported under 
NSF 9705131 and through the STCS under NSF 9120000. 
ISY was also supported by KOSEF 95-0702-03-01-3.
SLC was supported by DOE under grant DEFG02-96ER45439.
ISY, BKC, and SIL were supported by  the  
Creative Research  Initiative Project  of the  Ministry  of Science  and 
Technology, 
Korea.
Ames Laboratory is operated by the U.S. Department of Energy by Iowa State 
University
under Contract No. W-7405-Eng-82.

\ \\
\ \\
$^*$ Current address: Department of Material Science and Technology, K-JIST, 
Kwangju, 500-712, Korea\\
\ \\

\newpage

\begin{figure}
\caption{
The Raman spectra of YNi$_2$B$_2$C (T$_c$ = 15.3 K, $\Delta$T$_c$ = 0.4 K)
in B$_{2g}$ and B$_{1g}$ scattering geometries at various temperatures. 
The insets show
$\omega_p$/$\omega_p$(0), the normalized frequencies of the peaks  as
functions of the reduced temperature $T/T_c$.   
Determining $\omega_p$(0) is explained in the text.
The dotted curves are  for  $\Delta$/$\Delta_0$ predicted by 
the BCS theory. 
The error in $\omega_p$ comes from the broadness of the peak feature especially
near T = T$_c$.
Uncertainty in the reduced temperature is smaller than the size of the symbols.
}
\label{Yang-Fig1}
\end{figure}

\begin{figure}
\caption{
Raman spectra in  A$_{1g}$, B$_{1g}$, and B$_{2g}$ geometries  
of YNi$_2$B$_2$C  (T$_c$ = 15.3 K, $\Delta$T$_c$ = 0.4 K) 
and LuNi$_2$B$_2$C  (T$_c$ = 15.7 K, $\Delta$T$_c$ = 0.4 K) in superconducting
states and in normal states.  
The A$_{1g}$ spectra were taken using left-circularly polarized light in both 
excitation
and detection (left-left geometry). 
The spectrum denoted by x was taken at a magnetic field of 6T.
The peaks at $\sim$ 200 cm$^{-1}$ in the  A$_{1g}$ spectra 
are leakage of the strong B$_{1g}$ phonon due to slight mis-alignment of the
polarizer and analyzer. 
}
\label{Yang-Fig2}
\end{figure}

\begin{figure}
\caption{
 a)
Raman spectra of superconducting YNi$_2$B$_2$C(B$^{10}$ isotope,  
T$_c$ = 15.3 K, $\Delta$T$_c$ = 0.4 K) in B$_{2g}$ (solid line) and 
B$_{1g}$ (dashed line) symmetries. Raman spectra of the flux phase
are shown as solid dots along with a fit (thin line).
b) Assuming that {\it all} the Raman intensity at 250 cm$^{-1}$ is from the flux 
phase , the ``intrinsic'' Raman spectra of B$_{2g}$ (solid line) and B$_{1g}$ 
(dashed line)
are obtained by subtracting the ``flux'' contribution. This would be under-
estimation of the true intrinsic
spectra.
}
\label{Yang-Fig3}
\end{figure}

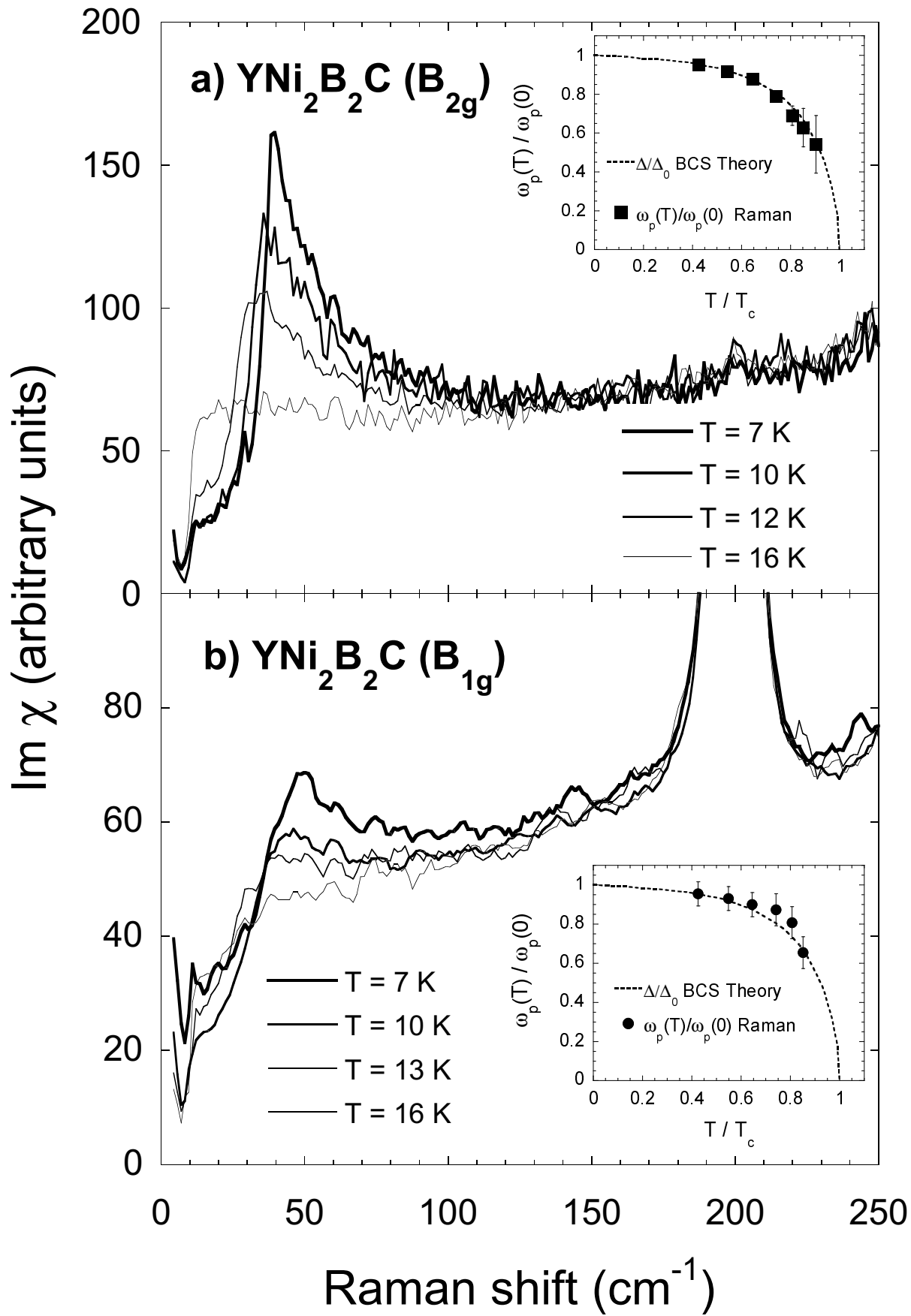
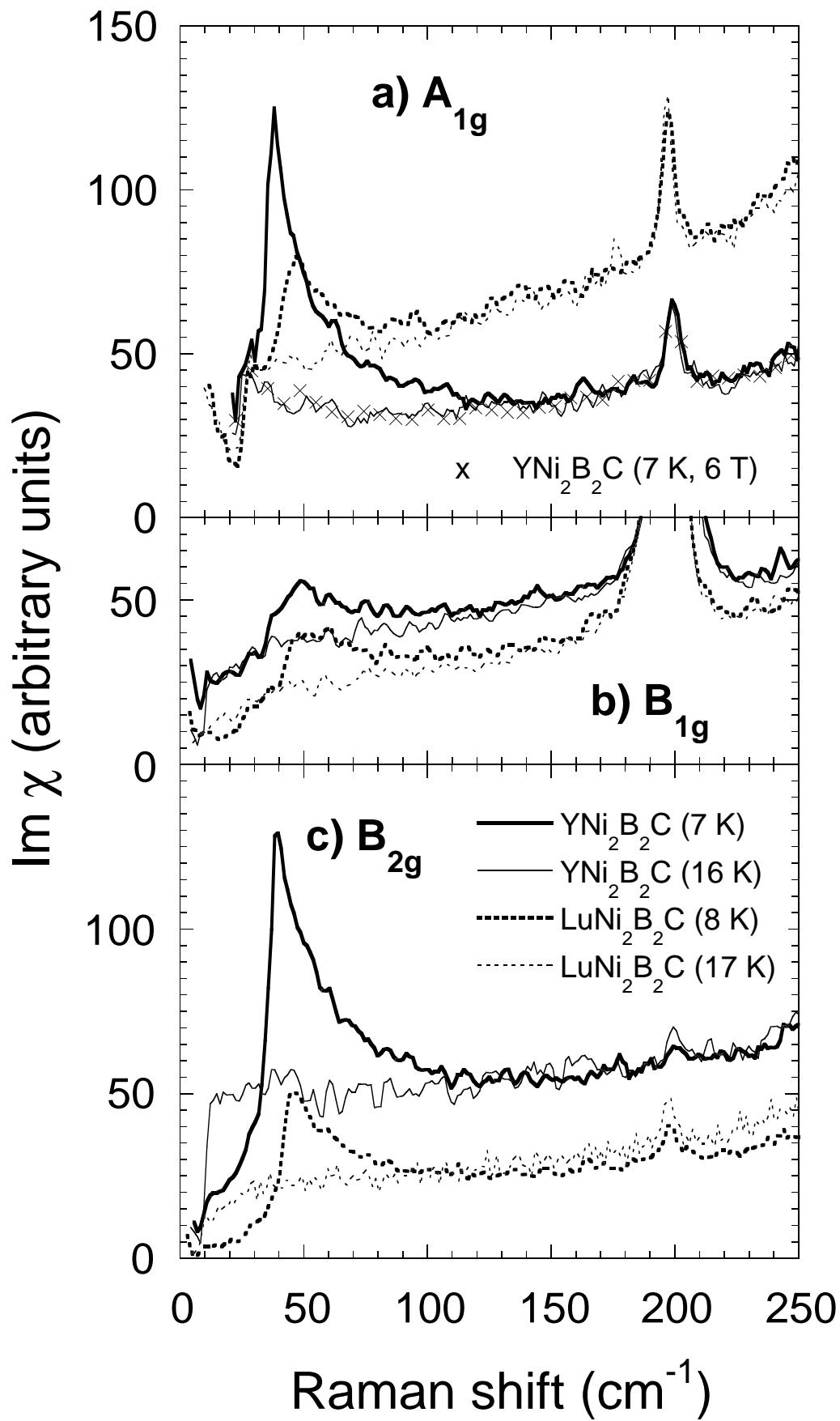
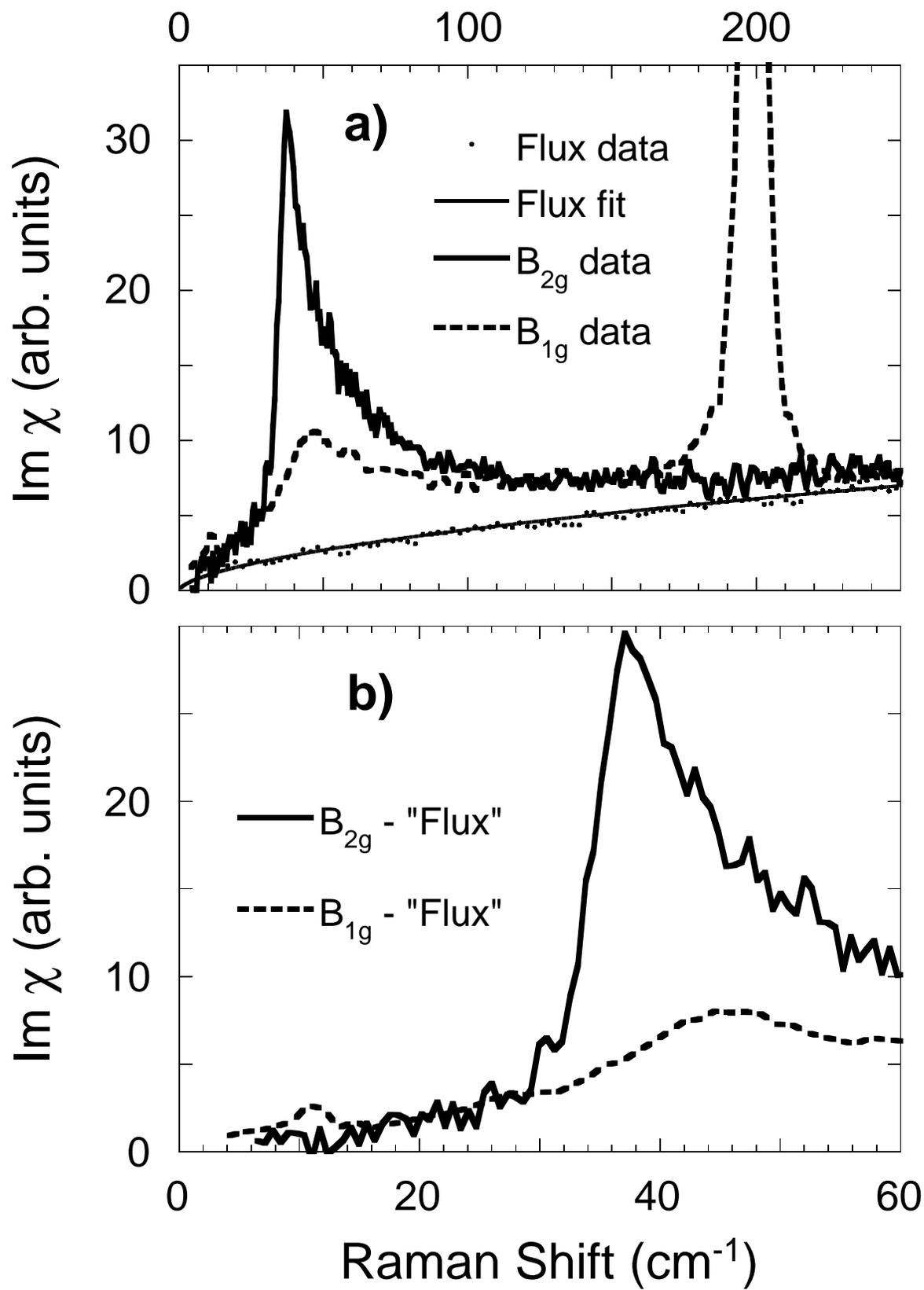
\end{document}